\documentstyle[12pt,psfig]{article}
\begin{document}
\begin{titlepage}
\begin{center}

{\large\bf{Analysis of  Low-\protect$x$  Gluon Density  from the
 $F_2$ Scaling Violations}}
\\[5.0ex]
{\large\it{ M. B. Gay  Ducati $^{*}$\footnotetext{${}^{*}$E-mail:gay@if.ufrgs.br}}}\\
 {\it and}\\
{ \large \it{ Victor P. B. Gon\c{c}alves $^{**}$\footnotetext{${}^{**}$E-mail:barros@if.ufrgs.br} 
}} \\[1.5ex]
{\it Instituto de F\'{\i}sica, Univ. Federal do Rio Grande do Sul}\\
{\it Caixa Postal 15051, 91501-970 Porto Alegre, RS, BRAZIL}\\[5.0ex]
\end{center}

{\large \bf Abstract:}
We present an analysis of the current methods of approximate determination of the gluon 
density $G(x,Q^2)$ at low-$x$ from the scaling violations of the proton structure 
function $F_2(x,Q^2)$. The gluon density is obtained from DGLAP equation by expansion
 at distinct points of expansion. We show that the different results given by the
proposed methods can be obtained from one new expression considering the adequate points of expansion. It is shown in which case the approximate determination of $G(x,Q^2)$
is reasonable.

\vspace{1.5cm}


\end{titlepage}

\section{Introduction}

Deep Inelastic Scattering (DIS)\cite{dis,4} of electrons on protons  
provides the classical test of the Quantum Chromodynamics (QCD)\cite{2,4} 
at short distances. A complete knowledge of the nucleon   
structure (parton picture of hadrons), including the 
gluon density, is fundamental for the design of future
proton collider experiments, but more immediate interest in the gluon
density centres on the main role it plays for physics at low $x$ and 
moderate $Q^2$. With the advent of HERA, we are working in a 
new kinematical 
region  both in $Q^2$ and  $x$. Whereas at large $Q^2$  the logarithmic scaling violations predicted by QCD  are expected, the direction 
towards small $x$ leads to a new kinematical region, where new dynamical effects 
 are expected to occur at a significant level. 

The standard perturbative QCD framework predicts that at a regime of low values 
of the Bjorken variable $x$ ($x\approx 10^{-4}$) and large values of $Q^2$, 
a nucleon consists predominantly of  the sea quarks and gluons. The gluons 
 couple only through   the strong interaction, consequently the gluons 
are not directly probed in the DIS, only contributing indirectly via the 
$g \rightarrow q\overline{q}$ transition. Therefore, its distribution is       
not so well determined as those of the quarks.

Recently, a direct determination of the gluon density was performed in the 
kinematical region $x>10^{-3}$ \cite{glh1}. A leading 
order determination of the gluon density in the proton was
performed by measuring multi-jet events from boson-gluon fusion
in deep-inelastic scattering with the detector H1 at the electron-proton
collider HERA. The  data agree with the dramatic rise of gluon density  
with decreasing $x$ predicted by the traditional analysis of  $F_2(x,Q^2)$ \cite{mrs,grv95}. These
 analysis have performed next-to-leading
order (NLO) QCD fits based on the Altarelli-Parisi (DGLAP)\cite{glap} 
evolution equations of the parton densities. The initial densities 
should be determined by the experiment at some scale 
$Q^2=Q_0^2$. The basic procedure is to parameterize the $x$ dependence 
at some low $Q_0^2$, but where perturbative QCD should be still applicable, and
then evolve up in $Q^2$ using NLO DGLAP equations to obtain the 
parton densities at all values of $x$ and 
$Q^2$ of the data. The input parameters 
are then determined by a global fit to the data. 

The parton densities are well determined for $x>0.5$, where there are 
many different types of high-precision constraints on the individual
parton distributions. The  exception is  the gluon density which is 
mainly constrained by (i) the momentum sum rule, (ii) prompt 
photon production and (iii) the scaling violations of $F_2$. On the other hand at small $x$ $(x\leq 10^{-3})$ we 
have only the measurement of  the structure function $F_2(x,Q^2)$, and many  
different partonic descriptions at small $x$, for example, DGLAP\cite{glap} and BFKL\cite{bfkl}. 
Therefore it is important to obtain more information in this region, mainly of the 
gluon density, which is the  dominant parton at small $x$. 

Using the fact that quark densities can be neglected and that  
the non-singlet contribution $F_2^{NS}$ can be ignored safely
at low $x$ in the DGLAP equations,  the equation for $F_2$  becomes, for
four flavours
\begin{eqnarray}
\frac{dF_2(x,Q^2)}{dlogQ^2} = \frac{10\alpha_s}{9\pi}\int_{x}^{1} 
dx'P_{qg}(x')\frac{x}{x'}g \left( \frac{x}{x'},Q^2 \right)\,\,\,,
\label{glap1}
\end{eqnarray}
where 
$g(x,Q^2)$ is the gluon density,
$xg(x,Q^2)=G(x,Q^2)$ is the gluon density momentum,
$\alpha_s=\alpha_s(Q^2)$ is the strong coupling constant and the  
splitting function
$P_{qg}(x')$ gives the probability of finding a quark inside a gluon  
with momentum $x'$ of the gluon.

In LO (leading order)
\begin{eqnarray}
P_{qg}(x')=\frac{1}{2}[x'^2+(1-x')^2]\,\,\,.
\end{eqnarray}

Substituting $x'=1-z$ we can write
\begin{eqnarray}
\frac{dF_2(x,Q^2)}{dlogQ^2} = \frac{10\alpha_s}{9\pi}\int_{0}^{1-x}
dzP_{qg}(1-z)G\left(\frac{x}{1-z},Q^2\right)\,\,\,.
\end{eqnarray}

Moreover, in  LO $P_{qg}(z)=P_{qg}(1-z)$, then
\begin{eqnarray}
\frac{dF_2(x,Q^2)}{dlogQ^2} = \frac{10\alpha_s}{9\pi}\int_{0}^{1-x}dz
P_{qg}(z)G\left(\frac{x}{1-z},Q^2\right)\,\,\,.
\label{dfglap}
\end{eqnarray}

This equation describes the evolution of the proton structure function 
$F_2(x,Q^2)$ in LO at small $x$. From experiments one can measure the 
structure function and then  determine its derivative 
$\frac{dF_2(x,Q^2)}{dlogQ^2}$. Some methods \cite{pritz,bora} were  
proposed in the literature to isolate the gluon density by its expansion inside  expression (\ref{dfglap}), 
but have different results for 
$G(x,Q^2)$. We demonstrate here that one of the methods is not totally correct and that, 
when it is corrected, the results of both methods  can be obtained by using  
different points of expansion from one  general expression. 
This article will be organized as follows. 
In  sect. 2 a 
revision of the  methods proposed in the literature is made. We 
show
that one of these methods is  not totally correct  and we present a modification. In  sect.
3 we propose a general expression for the approximate  
determination of the gluon density
   and determine the appropriate points
of expansion using data for 
 $\frac{dF_2(x,Q^2)}{dlogQ^2}$, obtained by H1\cite{dfh1}, Zeus\cite{dfzeus} and 
also for a screnning
model \cite{ayala} recently appeared in the literature. Finally, in the sect. 4, we present our conclusions.

\section{Analysis of Approximative Methods} 

 The  methods of approximate determination of the gluon density are 
based on the simplification
 of the convolution $P_{qg} \bigotimes g$  (eq. (\ref{glap1})) by the expansion of the
gluon density. The result is the gluon density $G(k.x)$ proportional to the derivative of
$F_2(x,Q^2)$, where the constant $k$ is associated to the choice of a point of expansion and $x=x_B$. In DIS we do not probe the gluon directly, but instead via the process $g \rightarrow q\overline{q}$. The longitudinal fraction $x_g$ of the proton's momentum that is 
carried by the gluon is therefore sampled over an interval bounded below by the 
Bjorken $x_B$ variable $x_B\equiv\frac{Q^2}{2p.q}$, where as usual $p$ and $q$ are 
the four momenta of the incoming proton and virtual photon respectively and
$Q^2\equiv-q^2$. Therefore, the choice of the point of expansion, and consequently of $k$,
is associated with the relation between $x_g$ and $x_B$. We discuss the  methods of approximate determination of gluon density proposed in the literature below.

\subsection{ Prytz's Approach}

Substituting  $P_{qg}(x')$  in the expression (\ref{dfglap}) and then expanding the gluon distribution around 
 $z=\frac{1}{2}$, one gets \cite{pritz}

\begin{eqnarray}
\frac{dF_2(x,Q^2)}{dlogQ^2} & \approx & \frac{5\alpha_s}{9\pi}
G(2x)\int_{0}^{1-x}dz[(1-z)^2+z^2]+ \nonumber \\
&   & +\frac{5\alpha_s}{9\pi}\frac{dG(z=\frac{1}{2})}{dz}
\int_{0}^{1-x}dz[(1-z)^2+z^2](z-\frac{1}{2})\,\,\,.
\label{df2priger}
\end{eqnarray}

Approximating the upper integration limit  $1-x \approx 1$ we can write

\begin{eqnarray}
\frac{dF_2(x,Q^2)}{dlogQ^2} & \approx & \frac{5\alpha_s}{9\pi}G(2x)\int_{0}^{1}dz[(1-z)^2+z^2]\\
& \approx & \frac{10\alpha_s}{27\pi}G(2x)\,\,\,. \label{dfpritz}
\end{eqnarray}

Therefore the gluon distribution can be expressed by 

\begin{eqnarray}
G(2x) & \approx & \frac {27\pi}{10\alpha_s}\frac{dF_2(x,Q^2)}{dlogQ^2}\,\,\,.
\label{gpritz}
\end{eqnarray}

The author of ref. \cite{pritz} claims that this relation is valid within 20\%  of accuracy, 
by taking into account many uncertainties: experimental measurement
of the $Q^2$-dependence of $F_2$, the value of  $\alpha_s$ and  the 
discarding of any typical
low-$x$ effects calculated beyond the DGLAP approximation, for example  
recombination effects\cite{100}. Estimates of gluon density \cite{glh1} have  demonstrated that this  approximation agrees with the gluon density obtained with other 
indirect determinations made by H1 and Zeus Collaborations.

\subsection{ Bora's {\it \bf et al.} Approach}

A different method is based in expansion of gluon distribution around $z=0$. Substituting the result in the 
expression (\ref{dfglap}) one gets \cite{bora}

\begin{eqnarray}
\frac{dF_2(x,Q^2)}{dlogQ^2}  \approx  \frac{5\alpha_s}{9\pi}
G(x)\left\{\frac{2(1-x)^3}{3} - (1-x)^2 +(1-x)\right\} + \nonumber \\
+\frac{5\alpha_s}{9\pi}\left\{\frac{(1-x)^2}{2} - \frac{2(1-x)^3}{3} + \frac{(1-x)^4}{2}\right\}\left\{x\frac{dG}{dx}\right\}\,\,\,.
\end{eqnarray}

Therefore

\begin{eqnarray}
\frac{dF_2(x,Q^2)}{dlogQ^2}  \approx \frac{5\alpha_s}{9\pi}A(x)\left\{
G(x)+\frac{B(x)}{A(x)}x\frac{dG}{dx}\right\}\,\,\,.\label{dfb}
\end{eqnarray}
where
\begin{eqnarray}
A(x)=\frac{2(1-x)^3}{3} - (1-x)^2 +(1-x)\,\,\,,
\end{eqnarray}
and
\begin{eqnarray}
B(x)=\frac{(1-x)^2}{2} - \frac{2(1-x)^3}{3} + \frac{(1-x)^4}{2}\,\,\,.
\end{eqnarray}

Bora {\sl et al.} \cite{bora} claim that this expression can be approximated by 

\begin{eqnarray}
\frac{dF_2(x,Q^2)}{dlogQ^2}  \approx \frac{5\alpha_s}{9\pi}\frac{[A(x)+B(x)]^2}
{A(x)+2B(x)}G\left(x + \frac{B(x)}{A(x)+B(x)}x\right)\label{dfbora}\\
\Rightarrow G\left(x + \frac{B(x)}{A(x)+B(x)}x\right) \approx \frac{9\pi}{5\alpha_s}
\frac{A(x)+2B(x)}{[A(x)+B(x)]^2}\frac{dF_2(x,Q^2)}{dlogQ^2} \,\,\,. 
\end{eqnarray}

In the limit $x\rightarrow0$ the last equation becomes

\begin{eqnarray}
G\left(\frac{4}{3}x\right) \approx \frac{9\pi}{5\alpha_s}
\frac{4}{3}\frac{dF_2(x,Q^2)}{dlogQ^2}\,\,\,.\label{gbora}
\end{eqnarray}

In figure  (\ref{fdfbpg}) we  compare the results of 
$\frac{dF_2(x,Q^2)}{dlogQ^2}$  for the approximate relations of  
Prytz (\ref{dfpritz}), Bora {\it et al.}(\ref{dfbora}) and the complete 
expression (\ref{dfglap}) using the gluon distribution GRV94(LO) \cite{grv95}.
In our calculations the value of $\alpha_s$ is obtained at $\Lambda_{\overline{MS}}^{(4)}=0.232 GeV^2$. The figure shows that Prytz approximation is more
accurate.

The result (\ref{gbora}) differs from Prytz result (\ref{gpritz}). Bora {\it et al.} explain that the difference 
arises because they  have retained the $x$ dependence in the upper limit 
of integration in  equation (\ref{dfglap}) after the expansion of the gluon density.
In order to show this is not the case, we have computed the expression  (\ref{df2priger}) without approximation in the upper integration limit. One gets 

\begin{eqnarray}
\frac{dF_2(x,Q^2)}{dlogQ^2} & \approx & \frac{5\alpha_s}{9\pi}[(1-x)-(1-x)^2+\frac{2}{3}(1-x)^3]\left\{
G(2x) + \frac{dG(z=\frac{1}{2})}{dz}\right.
\nonumber\\  
&  &\left.\left[ \frac{\frac{1}{2}(1-x)^2 -\frac{2}{3}(1-x)^3 +\frac{1}{2}(1-x)^4}{(1-x)-(1-x)^2+\frac{2}{3}(1-x)^3}
- \frac{1}{2} \right]\right\}\,\,\,.
\end{eqnarray}

Consequently,

\begin{eqnarray}
\frac{dF_2(x,Q^2)}{dlogQ^2} & \approx & \frac{5\alpha_s}{9\pi} [(1-x)-(1-x)^2+\frac{2}{3}(1-x)^3].\nonumber\\
&  & .G\left(x+\frac{\frac{1}{2}(1-x) -\frac{2}{3}(1-x)^2 +\frac{1}{2}(1-x)^3}{1-(1-x)+\frac{2}{3}(1-x)^2}x\right)\,\,\,.
\end{eqnarray}

In the limit $x\rightarrow 0$, this equation becomes

\begin{eqnarray}
\frac{dF_2(x,Q^2)}{dlogQ^2} & \approx & \frac{5\alpha_s}{9\pi}\frac{2}{3}G(2x)\\
\Rightarrow G(2x)& \approx & \frac{9\pi}{5\alpha_s}\frac{dF_2(x,Q^2)}{dlogQ^2}\,\,\,.
\end{eqnarray}

This result is identical to the Prytz one, demonstrating that Bora's argument is not correct.

 The Bora's method requires that 

\begin{eqnarray}
\frac{[B(x)]^2}{A(x)[A(x)+2B(x)]}\approx 0\,\,\,, \\
\frac{A(x)+B(x)}{A(x)+2B(x)} \approx 1\,\,\,.
\end{eqnarray}

In figure  (\ref{fraz}) we present the behaviour of these expressions at small $x$.   We immediately see that for small $x$ these approximations are not valid. 
Consequently the Bora approximation is not valid in this  $x$ range.

The more adequate approximation 
 of expression (\ref{dfb}) is
\begin{eqnarray} 
\frac{dF_2(x,Q^2)}{dlogQ^2} & \approx  & \frac{5\alpha_s}{9\pi}A(x)G\left(x + \frac{B(x)}{A(x)}x\right)\,\,\,.
\end{eqnarray} 

In the limit   $x\rightarrow 0$ , the equation becomes

\begin{eqnarray}
\frac{dF_2(x,Q^2)}{dlogQ^2} & \approx & \frac{5\alpha_s}{9\pi}\frac{2}{3}G\left(\frac{3}{2}x\right)\,\,\,.
\label{boracor}
\end{eqnarray}

In  figure  (\ref{fdfbbcpg}) we presented the behaviour of  
Prytz (\ref{dfpritz}) and Bora (\ref{dfbora}) results and one obtained from 
 expression (\ref{boracor}), 
called  Bora corrected. The expression (\ref{boracor}) also differs from  the 
expression of Prytz (\ref{dfpritz}). This is due to the election of a different point of 
expansion of the gluon density in both cases. In the next section we demonstrate that both 
results can be obtained from a general equation by choosing the  adequate  point of 
expansion.

\section{Expansion at an arbitrary point of expansion}

Using the expansion of the  gluon distribution $G\left(\frac{x}{1-z}\right)$ at an arbitrary
 $z=\alpha$  and retaining terms only up to the first derivative 
in the expansion, we get 

\begin{eqnarray}
\frac{dF_2(x,Q^2)}{dlogQ^2} \approx  \frac{5\alpha_s}{9\pi}\left\{\frac{2}{3}(1-x)^3-(1-x)^2+(1-x)\right\}.\nonumber\\
 .G\left[\frac{x}{1-\alpha}\left(1-\alpha+\frac{\frac{1}{2}(1-x)^3 
 -\frac{2}{3}(1-x)^2 + \frac{1}{2}(1-x)}{\frac{2}{3}(1-x)^2 
 -(1-x)+1}\right)\right]\,\,\,.\label{dfger}
\end{eqnarray}

In the limit   $x\rightarrow 0$ , the equation becomes

\begin{eqnarray}
\frac{dF_2(x,Q^2)}{dlogQ^2} & \approx & \frac{5\alpha_s}{9\pi}\frac{2}{3}
G\left[\frac{x}{1-\alpha}\left(\frac{3}{2} - \alpha\right)\right]\,\,\,.
\end{eqnarray}

When the points  $\alpha=\frac{1}{2}$ and  $\alpha=0$ are used, we get

\begin{eqnarray}
\frac{dF_2(x,Q^2)}{dlogQ^2} & \approx & \frac{5\alpha_s}{9\pi}\frac{2}{3}G(2x)\,\,\,,\\
\frac{dF_2(x,Q^2)}{dlogQ^2} & \approx & \frac{5\alpha_s}{9\pi}\frac{2}{3}
G\left(\frac{3}{2}x\right)\,\,\,.
\end{eqnarray}
Therefore the equation (\ref{dfger}) can be  reduced to  
Prytz  (\ref{dfpritz}) and  Bora corrected (\ref{boracor}), in their respective points of expansion. 

 The gluon distribution for a point of expansion $\alpha <1$
can be expressed by

\begin{eqnarray}
G\left[\frac{x}{1-\alpha}\left(\frac{3}{2} - \alpha\right)\right] & \approx & 
\frac{9\pi}{5\alpha_s}\frac{3}{2}\frac{dF_2(x,Q^2)}{dlogQ^2} \,\,\,.
\label{gger}
\end{eqnarray}

In  figure  (\ref{fexp2}) we present the results of $\frac{dF_2(x,Q^2)}{dlogQ^2}$ at some points of expansion $\alpha$ using the gluon 
distribution GRV94(LO), which fits better  the recent data of 
H1 Collaboration \cite{hera96}. At the current data of 
 $\frac{dF_2(x,Q^2)}{dlogQ^2}$ the points  $\alpha \ge 0.5$ are favoured. This 
means that in this kinematical region the  longitudinal momentum of the  gluon $x_g$ is more than twice the value of the longitudinal momentum of the probed quark (or antiquark) in DIS.

One can ask if  the approximative determination of the gluon 
density from $F_2(x,Q^2)$ scaling violations  can indicate
the presence of new effects at low-$x$, for example, recombination.
In figure (\ref{gscren}) we present the behaviour obtained using 
(\ref{gger}) 
for distinct points of expansion and  $\frac{dF_2(x,Q^2)}{dlogQ^2}$ data   calculated
 using the Glauber(Mueller) 
approach\cite{ayala}. This approach includes shadowing corrections   
and   $\frac{dF_2(x,Q^2)}{dlogQ^2}$ can be calculated  using the total cross section of
the gluon pair with the nucleon in the eikonal approach.

 We 
can conclude that, when compared with the behaviour of GRV94(LO), the 
screnning
can be cancealed for some values of $\alpha$ and that the better 
choices 
(with greater sensitivity to screnning) are in the range 
$0.5\leq \alpha < 0.8$.  This conclusion agrees with that of 
Ryskin {\sl et al.} \cite{ryskin} that estimate the value of the longitudinal 
gluon momenta $x_g$ in approximately 
3 times larger than the Bjorken $x_B$,
that corresponds to the expansion at $\alpha=0.75$.

\section{Conclusions}

The gluon  is by far the dominant parton in the small $x$ regime. Its 
distribution is not well determined as those of the quark, and there are several methods proposed in  the 
literature in order to  determine it 
 in different regions of $x$. 
More recently, new methods \cite{pritz,bora,cooper,charm} also 
based on the dominant behaviour  of the gluon distribution at small-$x$ were 
proposed.

In this work we discuss the approximative determination of the gluon density 
at low-$x$ from  $F_2(x,Q^2)$  scaling violations \cite{pritz,bora}.
We demonstrate that one of these methods is not correct. Moreover we proposed one 
general expression for approximative determination of the gluon density at 
arbitrary point of expansion of  $ G\left(\frac{x}{1-z}\right)$. Using    
data of H1 and Zeus Collaborations \cite{dfh1,dfzeus} and results obtained from
Glauber (Mueller) approach \cite{ayala} that includes shadowing corrections, we conclude that the more suitable points of expansion
(with greater sensitivity to screnning) are in the range $0.5\leq \alpha < 0.8$.
This conclusion agrees with recent results obtained by Ryskin {\sl et al.}\cite{ryskin}. 

The approximative determination of the gluon density obtained in this work was developed in 
the DGLAP framework. Consequently the approximation is valid in the DGLAP limit, which surely
breaks down at sufficiently small $x$. When   $\alpha_sln\frac{1}{x} \approx 1$ we should resum also the $\alpha_sln\frac{1}{x}$ contributions. In LO, $( \alpha_sln\frac{1}{x})^n$ resummation is
acomplished by the BFKL equation \cite{bfkl}. Therefore, a rigorous determination of the  transition region  to small-$x$ dynamics is very important in order to isolate correctly the gluon distribution. 

%The principal modification is the ressum of $\alpha_sln\frac{1}{x}$ terms using %$k_T$-factorization theorem,
%which modificate the splitting function $P_{qg}$, invalidating our work. %However, it has
%been pointed out \cite{ryskin} that such procedure masks the true dependence on %contributions from 
%the infrared region. It seems that, if DGLAP is invality in the actual region of data,  there is no %alternative but to work with the unintegrated gluon 
%distribution from BFKL and the  $k_T$-factorization theorem.

\section*{Acknowledgments}

 MBGD acknowledges C. A. Garcia Canal for enlightening discussions.
This work was partially financed by CNPq, BRAZIL.

\newpage
\section*{Figure Captions}

\vspace{1.0cm}

\begin{figure}
\centerline{\psfig{file=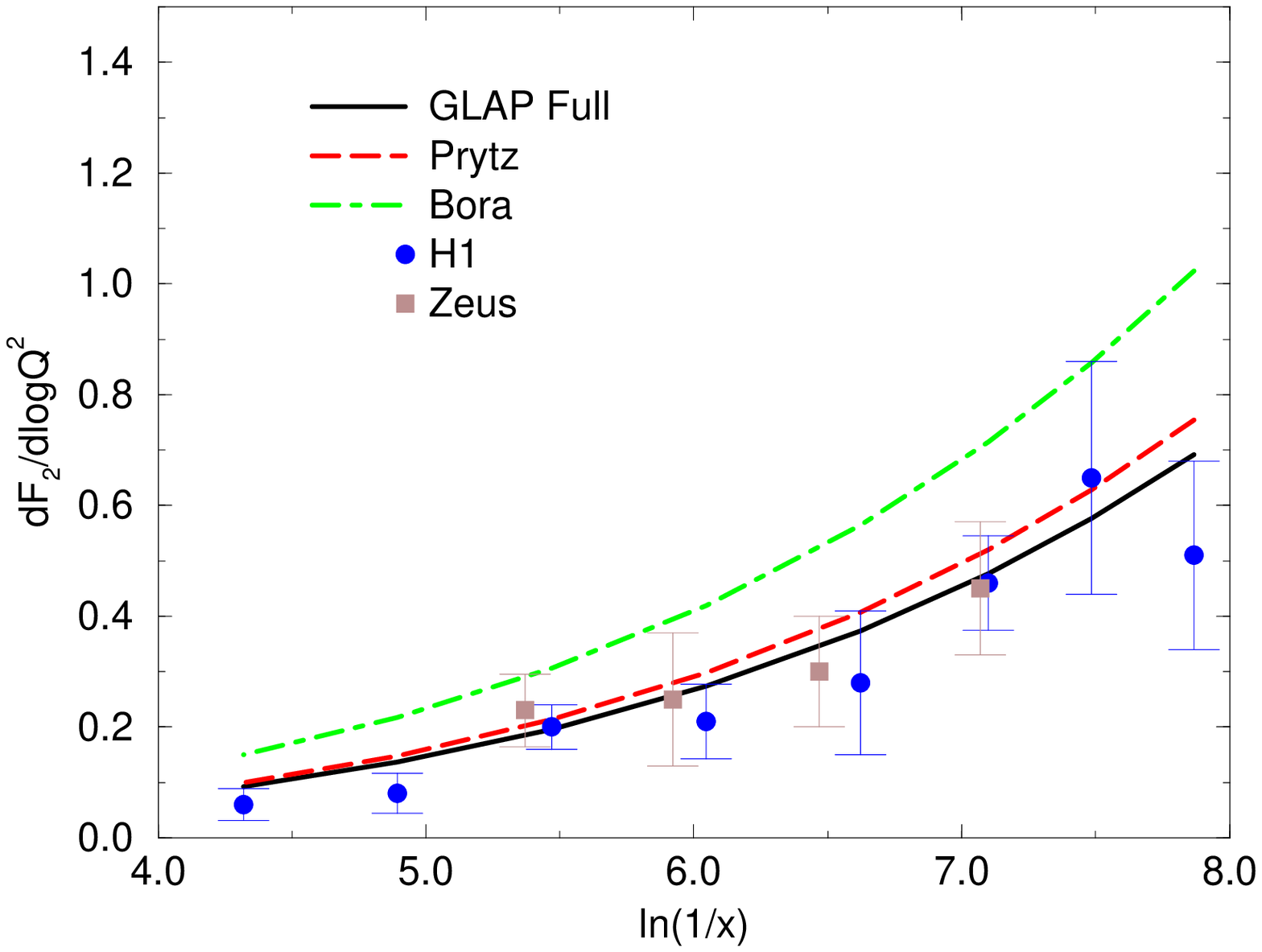,width=150mm}}
\caption{}
\label{fdfbpg}
\end{figure}

Fig. \ref{fdfbpg}:
Results of the  \protect$\frac{dF_2(x,Q^2)}{dlogQ^2}$ obtained 
from relations of Prytz (\protect\ref{dfpritz}), Bora (\protect\ref{dfbora}) 
and (\protect\ref{dfglap}),  using GRV94(LO)\protect\cite{grv95}. 
Data from
H1 \protect\cite{dfh1} and Zeus\protect\cite{dfzeus} at 
\protect$Q^2 = 20$ $GeV^2$. 

\begin{figure}
\centerline{\psfig{file=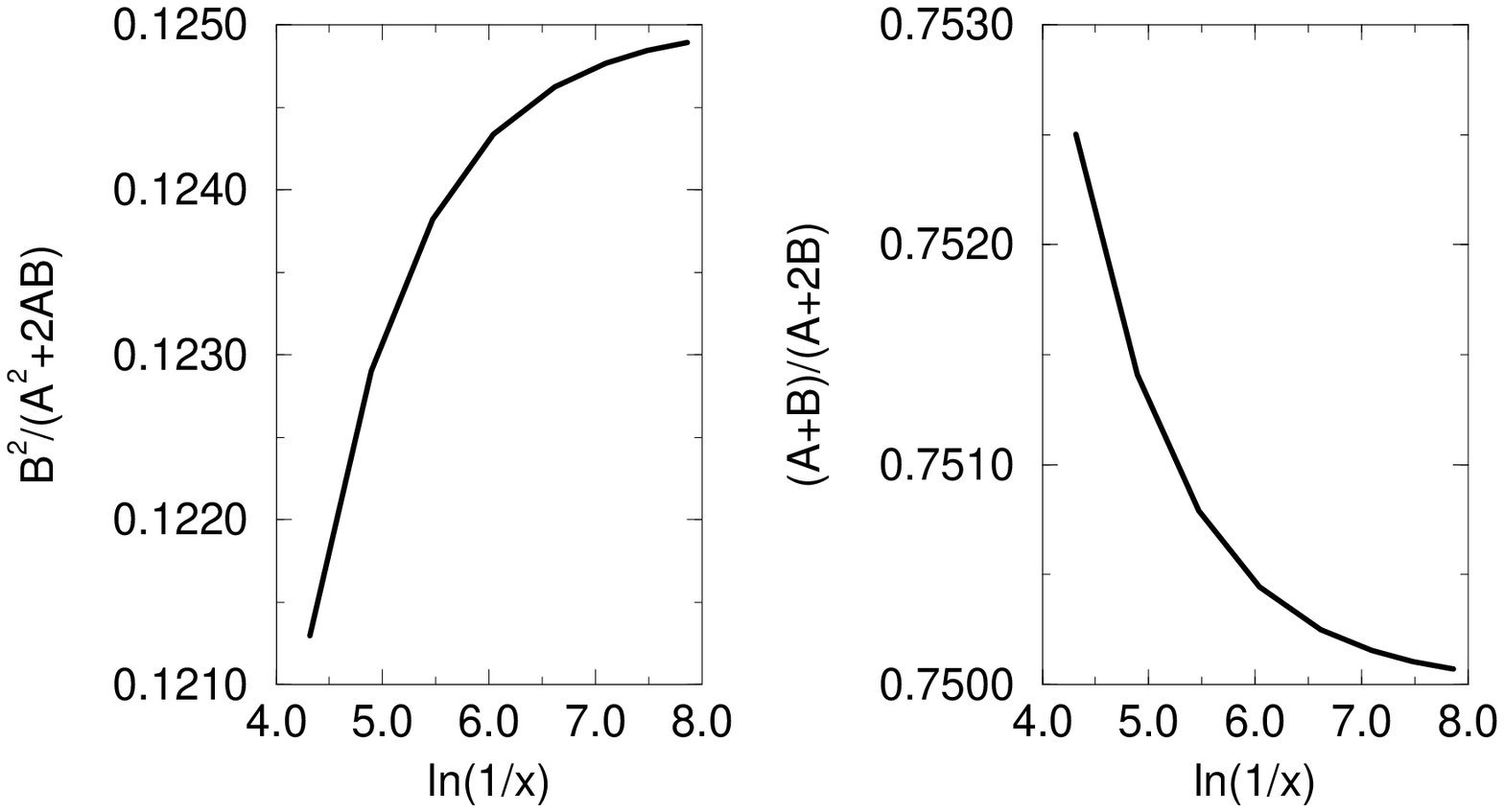,width=150mm}}
\caption{}
\label{fraz}
\end{figure}

\vspace{1.0cm}
Fig. \ref{fraz}:
 Behaviour of approximations used by Bora {\sl et al.} \cite{bora}.

\vspace{1.0cm}

\begin{figure}
\centerline{\psfig{file=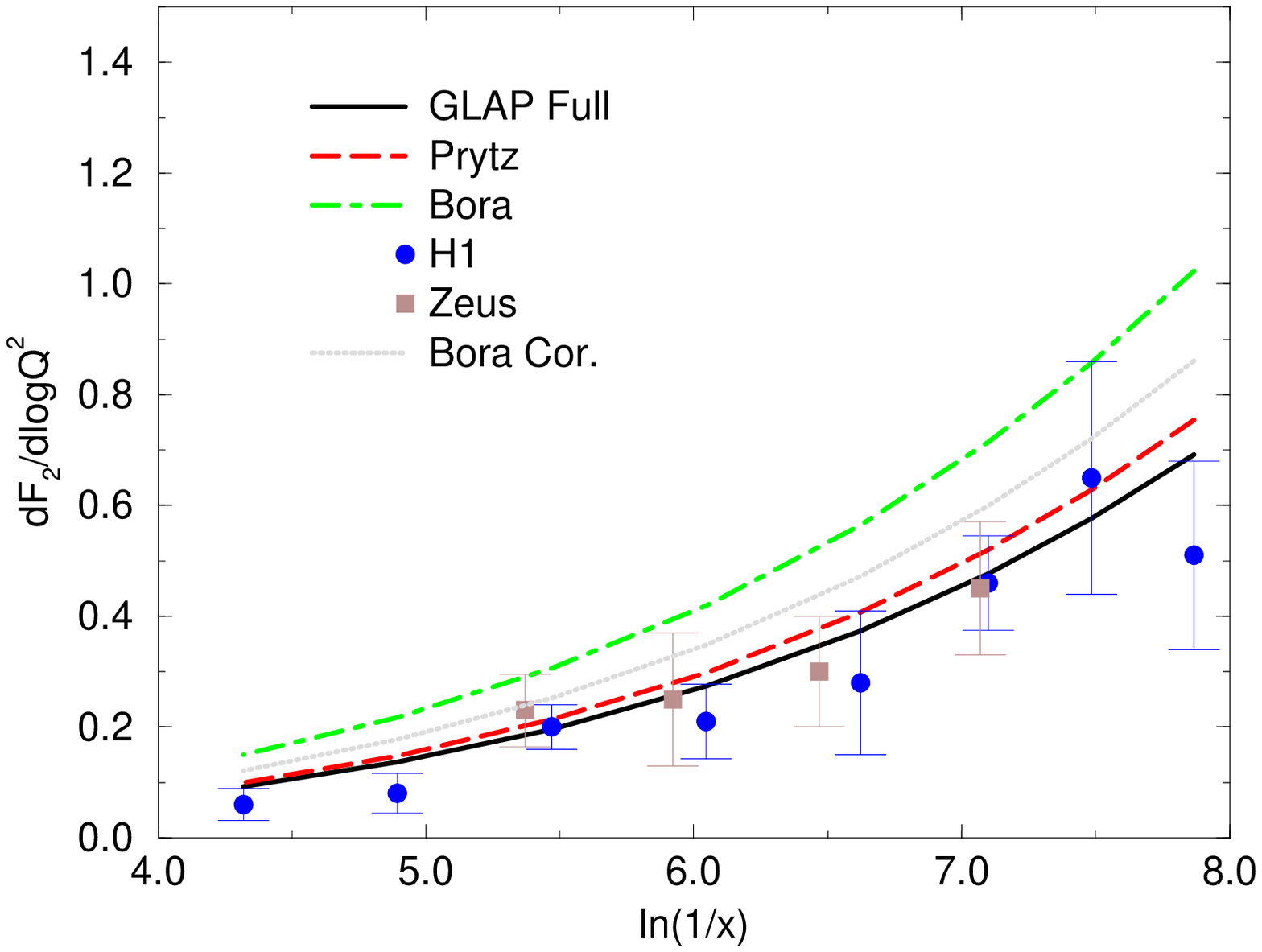,width=150mm}}
\caption{}
\label{fdfbbcpg}
\end{figure}

Fig. \ref{fdfbbcpg}:
Same as for  figure (\protect\ref{fdfbpg}), including results of 
equation  (\protect\ref{boracor})  presented for \protect$Q^2 = 20$ $GeV^2$.

\vspace{1.0cm}

\begin{figure}
\centerline{\psfig{file=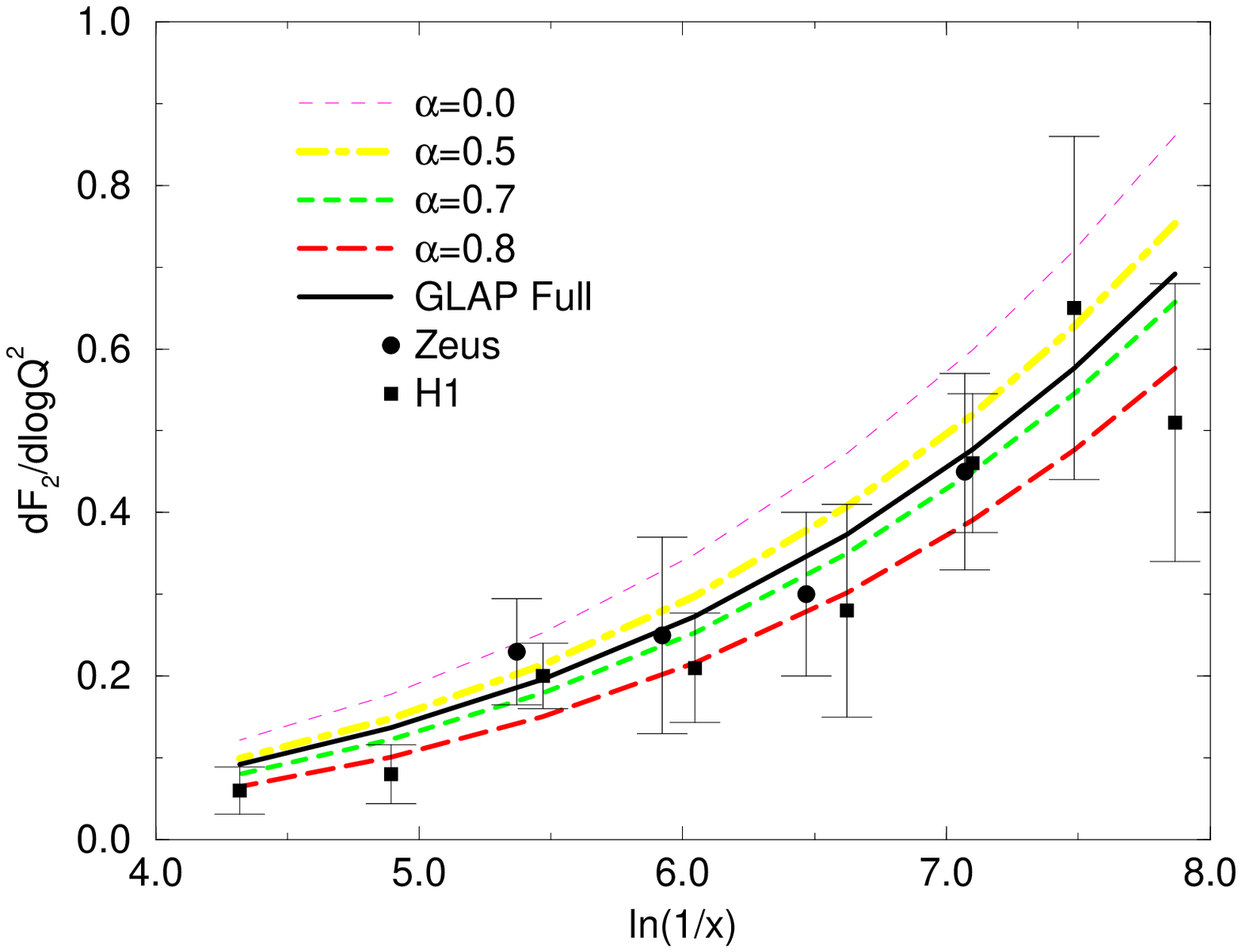,width=150mm}}
\caption{}
\label{fexp2}
\end{figure}

Fig. \ref{fexp2}:
Behaviour of  \protect$\frac{dF_2(x,Q^2)}{dlogQ^2}$ at several points of 
expansion $\alpha$. Also plotted the behaviour obtained using GRV94(LO) and 
the expression (\protect\ref{dfglap}) at \protect$Q^2 = 20$ $GeV^2$.

\vspace{1.0cm}

\begin{figure}
\centerline{\psfig{file=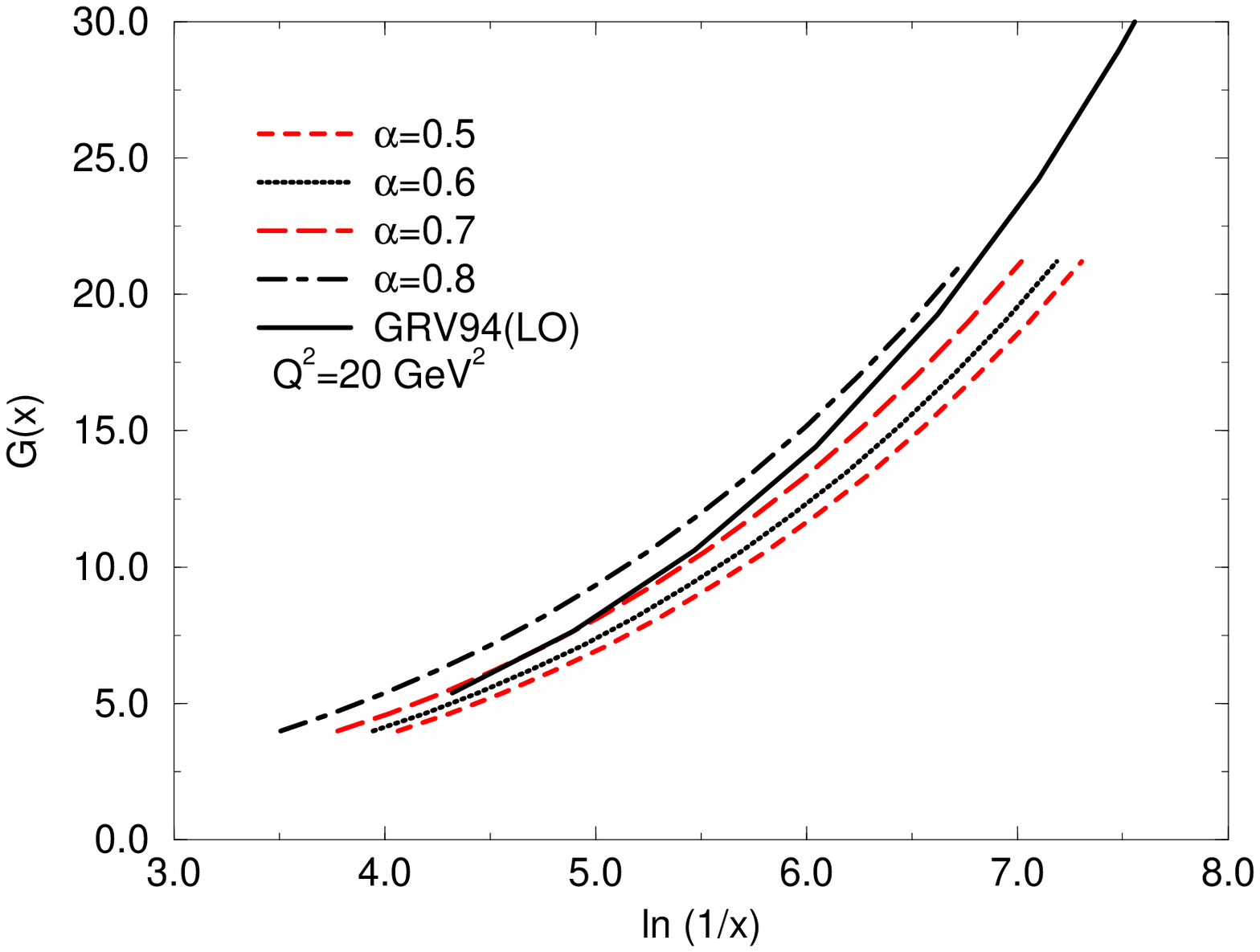,width=150mm}}
\caption{}
\label{gscren}
\end{figure}

Fig. \ref{gscren}:
Gluon distributions obtained of  (\protect\ref{gger}) using  
$\frac{dF_2(x,Q^2)}{dlogQ^2}$ with screnning\protect\cite{ayala}. 
Also shown the gluon distribution of 
GRV94(LO).

\newpage

\end{document}